\documentclass[twocolumn,showpacs,aps,prb]{revtex4}
\usepackage{amsmath}
\usepackage{amssymb}
\usepackage{bm}
\usepackage{color}
\usepackage{graphicx}

\begin{document}

\title{Non-linear frequency and amplitude modulation of a nano-contact spin torque oscillator}

\author{P.~K.~Muduli}
\email{muduli@kth.se} \affiliation{Materials Physics, Royal
Institute of Technology, Electrum 229, 164 40 Kista, Sweden}

\author{Ye.~Pogoryelov}
\affiliation{Materials Physics, Royal Institute of Technology,
Electrum 229, 164 40 Kista, Sweden}

\author{S.~Bonetti}
\affiliation{Materials Physics, Royal Institute of Technology,
Electrum 229, 164 40 Kista, Sweden}

\author{G.~Consolo}
\affiliation{Department of Physics, University of Ferrara, 44100 Ferrara, Italy}

\author{Fred~Mancoff}
\affiliation{Everspin Technologies, Inc., 1300 N. Alma School Road, Chandler, 85224, Arizona, USA}

\author{Johan~\AA kerman}
\affiliation{Materials Physics, Royal Institute of Technology,
Electrum 229, 164 40 Kista, Sweden}

\affiliation{Physics Department, University of Gothenburg, 41296 Gothenburg, Sweden}

\date{\today}
\begin{abstract}

We study the current controlled modulation of a nano-contact spin torque oscillator. Three
principally different cases of frequency non-linearity ($d^{2}f/dI^{2}_{dc}$ being zero, positive, and negative)
are investigated. Standard non-linear frequency
modulation theory is able to accurately describe the frequency shifts during modulation. However,
the power of the modulated sidebands only agrees with calculations based on a recent theory of
combined non-linear frequency and amplitude modulation.
\end{abstract}

 \maketitle

Spin-torque oscillators (STO) offer a combination of attractive properties such as 
ultra wide band frequency operation,~\cite{rippard2004prb,bonetti2009apl} extremely 
small footprint (without any need for large inductors), and easy integration using well-established magnetoresistive random access memory processes. The basic principle of a
spin-torque oscillator is based on the transfer of angular momentum
from a spin-polarized current to the local
magnetization.~\cite{slonczewski1996jmmm,berger1996prb} The effect
usually occurs in a nanoscale device where a large current density
($\sim10^{8}$ A/cm$^{2}$) can drive the precession of the free layer magnetization at
GHz frequencies,~\cite{tsoi2000nt,kiselev2003nt} thus acting as a
nanoscale oscillator. Effective modulation of the microwave signal generated from STOs is required 
for communication applications. However, both the STO frequency and amplitude are typically non-linear functions of the drive current.
This non-linearity is related to a change in the precession angle with the increase in the current magnitude.~\cite{houssameddine2007ntm,slavin2005ieeem,slavin2009ieeem}
Experiments have shown other sources of non-linearities such as
temperature~\cite{petit2007prl} and dynamic-mode hopping.~\cite{sankey2005prb,krivorotov2007prb,krivorotov2008prb} The wide range of possible sources of non-linear behavior is likely to render the frequency modulation of STOs highly non-trivial.

Despite the rapidly growing literature on the many different aspects of STOs, experimental studies of frequency modulation are still limited to a single work by Pufall \textit{et al.}~\cite{pufall2005apl} They observed both unequal sideband amplitudes and a shift of the carrier frequency with modulation amplitude, which they ascribed to non-linear frequency modulation (NFM). While linear frequency-modulation (LFM) theory assumes that the instantaneous frequency  of the modulated signal is linearly proportional to the modulating signal,~\cite{haykinbook} NFM theory takes into account the non-linear change in the intrinsic operating frequency during modulation. Pufall \textit{et al.} calculated the observed sideband amplitudes using NFM theory and found a rather large (about 50\%) discrepancy between their calculated and experimentally observed sidebands, which they argued might be due to amplitude modulation or other non-linear properties of the STO.

In this work we study the frequency and amplitude modulation of a nano-contact STO for various amounts of frequency non-linearity. The frequency non-linearity is described by the second derivative of the frequency, $f$, with respect to the dc bias current, $I_{dc}$ , $d^{2}f/dI^{2}_{dc}$. Three different cases of frequency non-linearity ($d^{2}f/dI^{2}_{dc}$ being zero, positive, and negative) are investigated. As expected from NFM theory, the carrier and its associated sidebands exhibit a change in frequency under modulation, which can be directly calculated from the experimentally determined non-linear properties of the frequency of the free-running STO. However, the \emph{power} of the modulated sidebands is only poorly reproduced using NFM theory and we show that it is essential to consider \textit{amplitude} modulation in order to reach any quantitative agreement. Using a recently proposed theory of combined nonlinear frequency and amplitude modulation (NFAM),~\cite{consolo2009arXiv} we are able to show remarkable agreement between our experimental data and calculations, which involve no adjustable parameters. Despite the complex phenomena involved in the STO non-linearities, we show that modulation of these devices is highly predictable.


The nano-contact metallic-based STOs studied in this work have been described in detail in Ref~\cite{mancoff2006apl}. Using e-beam lithography, a circular Al nano-contact with nominal diameter
of 130~nm is fabricated through a SiO$_2$ insulating layer, onto a 8$\times$26~$\mu$m$^{2}$ pseudo-spin-valve mesa with the following layer structure: Si/SiO$_2$/Cu(25 nm)/Co$_{81}$Fe$_{19}$(20 nm)/Cu(6 nm)/Ni$_{80}$Fe$_{20}$(4.5 nm)/Cu(3 nm)/Pd(2 nm). While all data presented here has been taken on a single device, similar behavior has been observed in several other devices of the same size.
\begin{figure}[t!]
\centering
\includegraphics[width=0.4\textwidth]{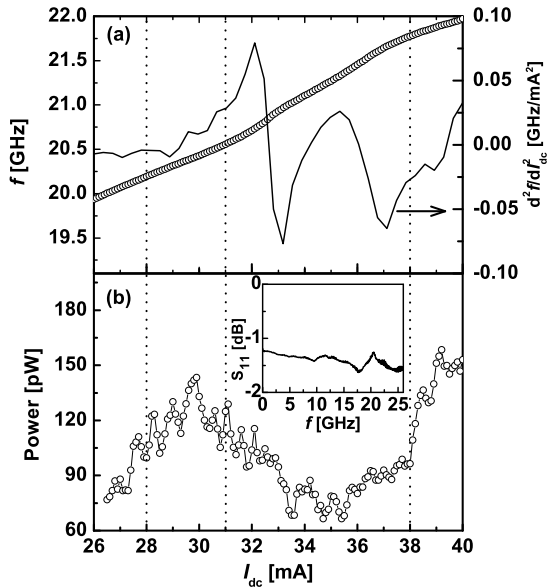}
\caption{Current dependence of the free running STO: (a)~frequency, $f$, and its second derivative, $d^{2}f/dI^{2}_{dc}; $ 
(b)~integrated power, both measured in a magnetic field of $H=$~10~kOe, applied at 
70~$^{\circ}$ to the film plane. Dotted lines indicate the three different operating points (28, 31, and 38~mA) used to compare  
three principally different 
cases of frequency non-linearity, corresponding to $d^{2}f/dI^{2}_{dc}$ being zero, positive, and negative, respectively. 
Inset in (b) shows the measured S-parameter, $S_{11}$ at the STO.}
\label{fig:cdep}
\end{figure}

The low frequency (100 MHz) modulating current is injected from an RF source to the STO via 
a circulator. The dc bias current is
fed to the device by a precision current source (Keithley 6221)
through a dc-40~GHz bias tee connected in parallel with the
transmission line. The signal is then amplified using a broadband 16-40~GHz,
+22~dB microwave amplifier, and finally detected by a spectrum
analyzer with an upper frequency limit of 46~GHz (Rohde \& Schwarz
FSU46). The actual RF current at the STO is calculated by taking into account 
losses and reflections due to impedance mismatch in the transmission line. Losses in our transmission 
line and circulator are characterized by injecting an input signal with the microwave source 
and measuring the output with the spectrum analyzer. The reflection at the STO is measured 
with a vector network analyzer and is shown in the  
inset of Fig.~\ref{fig:cdep}(b). 
The scattering matrix element $S_{11}$ shown in the figure is proportional 
to the amount of reflection at the STO, which is as high as $70-80$~\% over 
the entire measured frequency range, 0.01-26 GHz. All other components 
in the transmission line, which have nominal 50~$\Omega$ impedance, give 
a relative negligible contribution to the total amount of reflected signal. The 
signal detected at the spectrum analyzer is finally corrected for the standing waves in the transmission line. All data shown in this work 
have been corrected in order to compensate for all these effects.

\begin{figure}[t!]
\centering
\includegraphics[width=0.5\textwidth]{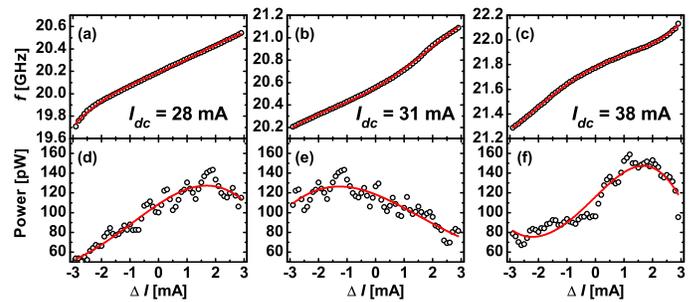}
\caption{(color online) Frequency and integrated power of the free-running STO around the dc bias current values of (a,d)~28~mA, (b,e)~31~mA and (c,f)~38~mA. The corresponding 4$^{th}$-order polynomial fits to frequency and the 3$^{rd}$-order polynomial fits to power are shown in solid red lines.}
\label{fig:cdepfit}
\end{figure}

The measurements are performed in a magnetic field of 10~kOe applied at an angle 
of 70$^{\circ}$ to the film plane to ensure that (\emph{i}) the STO operates around its maximum output power~\cite{bonetti2009apl}, and (\emph{ii})
only the so-called propagating mode~\cite{slonczewski1999jmmm, slavin2005prl, bonetti2009prl} is excited. This mode has a higher frequency than the ferromagnetic resonance mode and shows a blue-shift with bias current as confirmed in Fig.~\ref{fig:cdep}(a). Figure~\ref{fig:cdep} also shows that both the operating frequency and the integrated output power (which is proportional to the actual precession amplitude of the STO) [Fig.~\ref{fig:cdep}(b)] are strongly non-linear functions of the dc bias current. This behavior is likely related to the excitation of closely spaced discrete dynamic modes as the bias current is increased.~\cite{sankey2005prb,krivorotov2007prb,krivorotov2008prb} 

\begin{figure}[t!]
\centering
\includegraphics[width=0.5\textwidth]{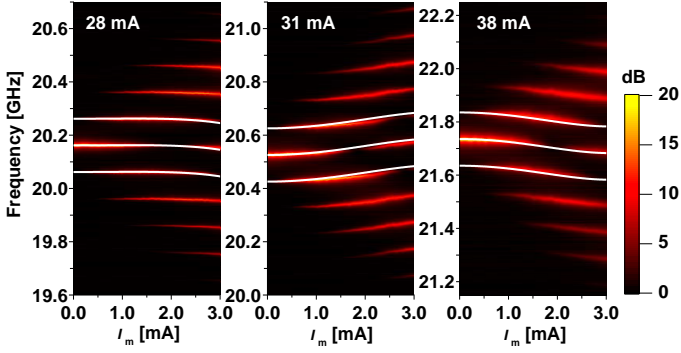}
\caption{(color online) Frequency modulation ($f_{m}=$~100~MHz) of the STO showing the progressive development of sidebands with increasing modulating amplitude $I_{m}$ at dc bias current values of 28~mA, 31~mA and 38~mA. Power is expressed in dB over the noise floor. The white lines show the calculated frequency of the carrier and the first-order sidebands according to the combined NFAM theory.}
\label{fig:fm}
\end{figure}

To test different non-linear modulation theories, we have chosen to focus on three principally different non-linear situations described by three different values of $d^{2}f/dI^{2}_{dc}$: zero,
positive, and negative, corresponding to a drive current of 28, 31, and 38~mA, respectively. These three operating points are shown as dotted lines 
in Fig.~\ref{fig:cdep}. The non-linearity can be more clearly seen in Fig.~\ref{fig:cdepfit}, which shows the frequency and integrated power of the free-running STO around these dc bias current values in a range equal to the maximum modulation current. The shape of frequency vs current at 28~mA is almost linear while it is convex for 31~mA and concave for 38~mA. The amplitude sensitivity is also clearly different at these current values, as seen from the corresponding plots of integrated power in Figs.~\ref{fig:cdepfit}(d)-2(f). Around these operating points we modulate the STO using a 100~MHz RF signal swept from 0 to 3~mA. The corresponding spectra are shown in Fig.~\ref{fig:fm} as a function of the modulating
current amplitude. In all three cases, the number of 
sidebands increases with increasing modulation amplitude. In the case of a linear frequency dependence (28~mA, $d^{2}f/dI^{2}_{dc}=0$) 
the carrier and sideband frequencies are entirely independent of the modulating
current (up to a modulation current of 2~mA). In contrast, both the carrier and the sideband frequencies show a clear
blue-shift at 31~mA and a clear red-shift at 38~mA as expected from the finite $d^{2}f/dI^{2}_{dc}$ with opposite signs. 

\begin{figure}[t!]
\centering
\includegraphics[width=0.45\textwidth]{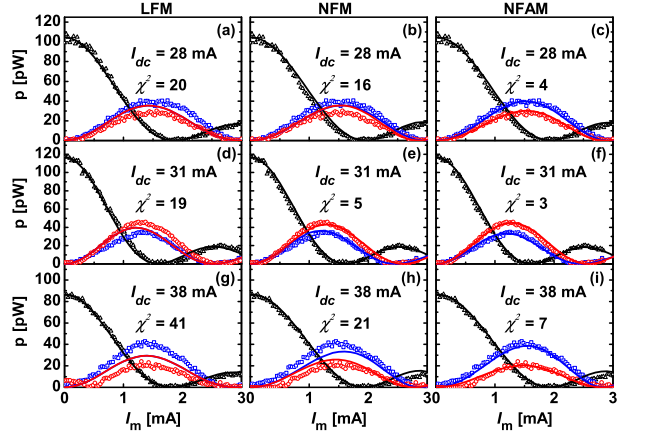}
\caption{(color online) Integrated power of the carrier (black triangles), and the first-order upper (blue squares) and lower (red circles) sidebands for the three different dc bias currents: row (a)-(c)~28~mA, row (d)-(f)~31 mA and row (g)-(h)~38~mA. First, 
second and third column shows the corresponding calculated integrated power (solid lines) as predicted 
by LFM, NFM and NFAM, respectively. The mean square error, $\chi^{2}$ between the experiment and calculated results of the two sidebands improved significantly for NFAM.}
\label{fig:sb}
\end{figure}

In Fig.~\ref{fig:sb}, we show the detailed modulation current dependence of the carrier and the first-order sideband power with calculated results as described in the following paragraph. While the evolution of the carrier power with modulation current does not seem to be affected by the non-linearity, both the upper and lower sidebands are strongly affected by the sign and the value of $d^{2}f/dI^{2}_{dc}$: the lower sideband gets markedly \textit{stronger} than the upper sideband for $d^{2}f/dI^{2}_{dc}>0$ (31~mA), and \textit{weaker} than the upper sideband for $d^{2}f/dI^{2}_{dc}<0$ (38~mA). The position of the maximum sideband power is also shifted up/down for the upper/lower sideband. It is noteworthy that this shift only depends on the magnitude of $d^{2}f/dI^{2}_{dc}$ and does not change sign when $d^{2}f/dI^{2}_{dc}$ goes from positive to negative. Even for the linear case (28~mA, $d^{2}f/dI^{2}_{dc}=0$), the power of the two sidebands are unequal. The upper sideband has higher power than the lower sideband, as expected from the positive slope of amplitude versus bias current in Fig.~\ref{fig:cdepfit}(d). This case of linear frequency modulation provides a strong experimental evidence that amplitude modulation is also taking place.

In order to interpret the observed behavior and estimate the importance of both the frequency and amplitude non-linearities, we consider three qualitatively different models describing (\textit{i})~LFM, (\textit{ii})~NFM, and (\textit{iii}) NFAM. The latter model is adapted from~\cite{consolo2009arXiv} and specifically takes into account non-linearities in both output frequency and amplitude as a function of the input bias current.

Since LFM and NFM models have already been described in Ref~\cite{pufall2005apl, haykinbook}, we focus on the details of the NFAM model used in our analysis. The instantaneous frequency  is assumed to depend nonlinearly on the modulating signal:
\begin{equation} \label{eq:fi}
	f_{i}(t)=k_{0}+k_{1} m(t)+ k_{2}m(t)^{2}+k_{3}m(t)^{3}+...,
\end{equation}
where, $m(t)$, is the modulating signal and the coefficients $k_{i}$ represent the $i$-th order frequency sensitivity coefficients. Similarly, the output amplitude, $A_{c}$ is given by
\begin{equation}\label{eq:ac}
	A_{c}(t)=\lambda_{0}+\lambda_{1} m(t)+ \lambda_{2}m(t)^{2}+\lambda_{3}m(t)^{3}+...,
\end{equation}
where $\lambda_{i}$ is \textit{i}th order amplitude sensitivity coefficient. The coefficients $k_{i}$ and $\lambda_{i}$ are given by the non-linear current dependence of $f$ and $A$ of the free running STO. We use sine wave modulation, $m(t) = I_{m}sin(2\pi f_{m}t)$, where $I_{m}$ is the amplitude and $f_{m}$ is the frequency of modulating signal. The resulting NFAM spectrum becomes~\cite{consolo2009arXiv}
\begin{equation}\label{eq:sp}
\begin{split}
	S(f)=&\frac{1}{4}\sum^{3}_{h=0} \gamma_{h} \sum^{\infty}_{n, m, p, q=-\infty}J_{n}(\beta_{1})J_{m}(\beta_{2})J_{p}(\beta_{3})J_{q}(\beta_{4})\\
 &\left[\delta(f-f^{I}_{c}-(n+2m+3p+4q+h)f_{m})\right.\\
 	&+\delta(f-f^{I}_{c}-(n+2m+3p+4q-h)f_{m})  \\
	&+\delta(f+f^{I}_{c}-(n+2m+3p+4q+h)f_{m}) \\
 	&+\left.\delta(f+f^{I}_{c}-(n+2m+3p+4q-h)f_{m})\right]\end{split}
\end{equation}
\begin{table*}[t]\footnotesize
\caption{Modulation sensitivity coefficients found from polynomial fits of frequency and amplitude of the free running STO.}\label{tab:ms}
\begin{tabular*}{1.0\textwidth}{@{\extracolsep{\fill}} @{}c@{}@{}c@{}@{}c@{}@{} c @{}@{} c @{}@{} c @{}@{} c @{}@{} c @{}@{} c @{}@{} c@{} }

\hline\hline
Current         & $k_{0}$ & $k_{1}$ & $k_{2}$ & $k_{3}$ & $k_{4}$ & $\lambda_{0}$ & $\lambda_{1}$ & $\lambda_{2}$ & $\lambda_{3}$\\[4pt]
\hline
(mA)						& (GHz) & (MHz/ & (MHz/ & (MHz/ & (MHz/ & (pW$^{1/2}$)& (pW$^{1/2}$/ & (pW$^{1/2}$/ & (pW$^{1/2}$/\\[2pt]
						&  & mA) & mA$^{2}$) & mA$^{3}$) & mA$^{4}$) &  & mA) & mA$^{2}$) & mA$^{3}$)\\[4pt]
\hline
28         & 20.185 & $117\pm 1$ & $1\pm 1$ & $2\pm 0.2$ & $8\pm 1$ & $10.4\pm 0.5$ & $0.9\pm 0.07$ & $-0.2\pm 0.02$ & $-0.03\pm 0.01$ \\
31         & 20.545  & $147\pm 1$ & $20\pm 1$ & $0.8\pm 0.1$ & $-1\pm 0.1$ & $10.9\pm 0.6$ & $-0.5\pm 0.07$ & $-0.15\pm 0.02$ & $0.02\pm 0.01$ \\
38         & 21.779 & $115\pm 1$ & $-22.5\pm 0.6$ & $-3.3\pm 0.1$ & $1.6\pm 0.1$ & $10.8\pm 1$ & $1.3\pm 0.07$ & $-0.1\pm 0.02$ & $-0.12\pm 0.01$ \\
\hline\hline
\end{tabular*}
\end{table*}
where $\beta_{1}=k_{1}I_{m}/f_{m}+3k_{3}I_{m}^{3}/4f_{m}$, $\beta_{2}=k_{2}I_{m}^{2}/4f_{m}+k_{4}I_{m}^{4}/4f_{m}$, $\beta_{3}=k_{3}I_{m}^{3}/12f_{m}$, and $\beta_{4}=k_{4}I_{m}^{4}/32f_{m}$ are frequency modulation indices of different order. $\gamma_{0}=\lambda_{0}+\lambda_{2} I_{m}^{2}/2$, $\gamma_{1}=\lambda_{1}I_{m}+3\lambda_{3} I_{m}^{3}/4$, $\gamma_{2}=\lambda_{2} I_{m}^{2}/2$, and $\gamma_{3}=\lambda_{3} I_{m}^{3}/4$ are amplitude modulation parameters. In the above we assumed that the frequency in Eq.~(\ref{eq:fi}) is non-linear up to fourth order and the amplitude in Eq.~(\ref{eq:ac}) is non-linear up to third order, which is found sufficient to describe the experimental data. The frequency spectrum $S(f)$ consists of a \textit{shifted} carrier frequency
\begin{equation} \label{eq:fc}
	f^{I}_{c}=k_{0}+k_{2}I_{m}^{2}+3k_{4} I_{m}^{4}/8+...
\end{equation}
and an infinite number of sidebands symmetrically located at $f^{I}_{c}\pm lf_{m}$, where $l=n+2m+3p+4q\pm h$ is a positive integer identifying 
the sideband order. The NFAM carrier shift is identical to that obtained from an NFM model since effects due to amplitude modulation do not enter in Eq.~(\ref{eq:fc}). This shift can be readily calculated by means of  the polynomial fitting procedure shown in Fig.~\ref{fig:cdepfit}. The comparison with the experimentally obtained values reveals a good agreement, as shown in Fig.~\ref{fig:fm}. The sideband power, on the other hand, is strongly affected by the amplitude modulation, through the
coefficients $\gamma_{i}$, and can be used to compare the NFM and NFAM models. 
In a 6 mA interval around each operating point, we expand the frequency dependence into a fourth-order Taylor series, and the amplitude dependence into a  third-order Taylor series as shown in Fig.~\ref{fig:cdepfit}. The coefficients along with their standard errors are summarized in Table~\ref{tab:ms}. Using these coefficients we calculate the sideband power expected from NFM and NFAM, respectively (second and third column in Fig.~\ref{fig:sb}) and also compare with LFM theory (first column in Fig.~\ref{fig:sb}).

LFM theory completely fails to describe the strong asymmetry between the upper and lower sidebands in all cases. In the linear case of 28~mA [Figs.~\ref{fig:sb}(a)-~\ref{fig:sb}(c)] both NFM and LFM theory predict nearly the same behavior with equal sideband power since only $k_{1}$ is significant and $k_{2}\approx 0$. In contrast, the NFAM model correctly produces both the upper and lower sideband power, implying a much better agreement, mostly captured by the amplitude modulation sensitivity coefficient $\lambda_{1}$. In fact, the mean square error, $\chi^{2}$ between the experiment and calculated results of the two sidebands decreases by about 80~\% for NFAM theory compared to LFM. In the two non-linear cases, the NFM model captures the change in sign of the sideband asymmetry, given by the sign change in $k_{2}$, but only yields a partial improvement compared to LFM. On the contrary, when the amplitude sensitivity coefficients are also taken into account 
the agreement of the calculations with experiment is essentially perfect. This agreement is only obtained when \textit{both} frequency and amplitude non-linearities are accounted for; both $k_{2}$ and $\lambda_{1}$ are significant. For 31 mA (38 mA), the mean square error between the experiment and calculated results of the two sidebands decreases by about 85~\% (83~\%) for NFAM theory compared to LFM and about 10~\% (36~\%) compared to NFM theory. We emphasize that none of the presented calculations involve \textit{any} free parameters and are completely determined by the experimentally measured nonlinear current dependences of the free-running STO. The agreement with NFAM was also found to be valid for a range of lower modulation frequencies (down to 40 MHz) over the entire range of dc bias currents. 
Thus our results show that, as long as both non-linearities are accounted for, the proposed scheme of combined modulation is able to accurately predict the resulting sideband powers and frequency shifts over a wide range of varying operating conditions. Consequently, the STO behaves as an ordinary RF oscillator and should lend itself to communication applications.

In conclusion, we have carried out a detailed modulation study on a nano-contact STO. In particular, we have studied the impact of different levels of frequency non-linearity. In the non-linear cases, both carrier and sidebands frequencies are shifted as a function of the modulation current. Both frequency and amplitude non-linearities produce a significant asymmetry in the power of the upper and lower sidebands. We find that a combined non-linear frequency and amplitude modulation model can accurately describe all our experimental data without any adjustable parameters. The modulation of an STO is therefore predictable and independent of the complex mechanism behind the non-linearity. The results are significant for the continued development of communication and signal processing applications of spin torque oscillators.

Support from the Swedish Foundation for Strategic Research (SSF),
the Swedish Research Council (VR), the G\"{o}ran Gustafsson Foundation and
the Knut and Alice Wallenberg Foundation are gratefully acknowledged. Johan~\AA kerman is a Royal Swedish Academy of Sciences Research Fellow supported by a grant from the Knut and Alice Wallenberg Foundation. Giancarlo Consolo gratefully thanks support from CNISM through "Progetto Innesco". We thank Randy K. Dumas for critical reading of the manuscript.

\end{document}